\RequirePackage{lineno} 
\documentclass[traditabstract,printer,longauth]{aa}
\usepackage{txfonts}
\usepackage[dvips]{graphics}
\usepackage[dvips]{graphicx}
\usepackage{graphicx}
\usepackage{longtable}
\usepackage{color}
\usepackage{url}
\usepackage{dcolumn} 

\usepackage{natbib}
\bibpunct{(}{)}{;}{a}{}{,} 

\newcommand{\SHBLc}{SHBL~J001355.9--185406}
\newcommand{\SHBL}{\SHBLc\ }
\newcommand{\dgr}{\ensuremath{^\circ}}
\newcommand{\ergcms}{\ensuremath{\mathrm{erg}\ \mathrm{cm}^{-2}\ \mathrm{s}^{-1}}}%

\newcommand{\ergs}{erg\,s$^{-1}$}%
\newcommand{\cms}{cm$^{-2}$\,s$^{-1}$}
\newcommand{\hess}{\textsc{H.E.S.S.}}

\newcommand{\swift}{\textsl{Swift}}
\newcommand{\swiftxrt}{\textsl{Swift}-XRT}
\newcommand{\swiftuvot}{\textsl{Swift}-UVOT}
\newcommand{\fer}{{\sl {\it Fermi}}}
\newcommand{\fla}{\fer-LAT}
 \newcommand{\fvar}{\mathrm{F}_{\mathrm{var}}}

\newcommand{\gr}{$\gamma$-ray}
\newcommand{\grs}{$\gamma$-rays}
\newcommand{\indexHESSnumOnly}{\ensuremath{\rm 3.4 \pm 0.5_{\rm{stat}} \pm 0.2_{\rm{sys}}}}
\newcommand{\indexHESS}{\ensuremath{\rm \Gamma = 3.4 \pm 0.5_{\rm{stat}} \pm 0.2_{\rm{sys}}}}
\newcommand{\hessfluxEdec}{\ensuremath{1.16 \pm 0.23_{\rm{stat}}\pm 0.23_{\rm{sys}}}}

\begin{document}


\modulolinenumbers[5]

\title{Discovery of high and very high-energy emission from the BL~Lac object \SHBL}
\author{H.E.S.S. Collaboration
\and A.~Abramowski \inst{1}
\and F.~Acero \inst{2}
\and F.~Aharonian \inst{3,4,5}
\and A.G.~Akhperjanian \inst{6,5}
\and E.~Ang\"uner \inst{7}
\and G.~Anton \inst{8}
\and S.~Balenderan \inst{9}
\and A.~Balzer \inst{10,11}
\and A.~Barnacka \inst{12}
% ...
}

\institute{
Universit\"at Hamburg, Institut f\"ur Experimentalphysik, Luruper Chaussee 149, D 22761 Hamburg, Germany \and
Laboratoire Univers et Particules de Montpellier, Universit\'e Montpellier 2, CNRS/IN2P3,  CC 72, Place Eug\`ene Bataillon, F-34095 Montpellier Cedex 5, France \and
Max-Planck-Institut f\"ur Kernphysik, P.O. Box 103980, D 69029 Heidelberg, Germany \and
Dublin Institute for Advanced Studies, 31 Fitzwilliam Place, Dublin 2, Ireland \and
National Academy of Sciences of the Republic of Armenia, Yerevan  \and
Yerevan Physics Institute, 2 Alikhanian Brothers St., 375036 Yerevan, Armenia \and
Institut f\"ur Physik, Humboldt-Universit\"at zu Berlin, Newtonstr. 15, D 12489 Berlin, Germany \and
Universit\"at Erlangen-N\"urnberg, Physikalisches Institut, Erwin-Rommel-Str. 1, D 91058 Erlangen, Germany \and
University of Durham, Department of Physics, South Road, Durham DH1 3LE, U.K. \and
DESY, D-15735 Zeuthen, Germany \and
Institut f\"ur Physik und Astronomie, Universit\"at Potsdam,  Karl-Liebknecht-Strasse 24/25, D 14476 Potsdam, Germany \and
Nicolaus Copernicus Astronomical Center, ul. Bartycka 18, 00-716 Warsaw, Poland
% ...
}

\authorrunning{\hess\ collaboration}
\titlerunning{Discovery of \SHBL}

\abstract{The detection of the high-frequency peaked BL~Lac object (HBL) \SHBL
($z$=0.095) at high (HE; 100~MeV$<$E$<$300~GeV) and very high-energy (VHE;
$E>100\,{\rm GeV}$) with the \fer\ Large Area Telescope (LAT) and the High
Energy Stereoscopic System (\hess) is reported. Dedicated observations were
performed with the \hess\ telescopes, leading to a detection at the $5.5\,\sigma$
significance level. The measured flux above 310~GeV is $(8.3 \pm
1.7_{\rm{stat}}\pm 1.7_{\rm{sys}})\times 10^{-13}$ photons \cms\ (about 0.6\% of that of
the Crab Nebula), and the power-law spectrum has a photon index of \indexHESS.
Using 3.5 years of publicly available \fla\ data, a faint counterpart has been
detected in the LAT data at the $5.5\,\sigma$ significance level, with an integrated
flux above 300~MeV of $(9.3 \pm 3.4_{\rm stat} \pm 0.8_{\rm
sys})\times 10^{-10}$ photons \cms\ and a photon index of $\Gamma = 1.96 \pm 0.20_{\rm
stat} \pm 0.08_{\rm sys}$. X-ray observations with \swiftxrt\ allow the synchrotron peak energy in $\nu F_\nu$ representation to be located at $\sim 1.0\,{\rm
keV}$. The broadband spectral energy distribution is modelled with a one-zone
synchrotron self-Compton (SSC) model and the optical data by a black-body
emission describing the thermal emission of the host galaxy. The derived parameters
are typical of HBLs detected at VHE, with a particle-dominated jet.
}

\offprints{\\David Sanchez - email: david.sanchez@mpi-hd.mpg.de
\\Jonathan Biteau - email: biteau@in2p3.fr}


\maketitle

\section{Introduction} \label{intro}

Observations of blazars in \grs\ offer the unique possibility to probe one of
the most violent phenomena in the Universe. Blazars are active galactic nuclei
(AGN) with their jets pointing towards the observer \citep{THEO:Unification2}.
Two classes have been distinguished \citep[see e.g. ][]{2012MNRAS.420.2899G,2011AIPC.1381..180G}: the flat spectrum radio quasars (FSRQ) are the
most powerful blazars and their optical spectra exhibit absorption and emission
lines, whereas the BL~Lacertae (BL~Lac) class is less luminous and presents
weaker emission lines, with the equivalent widths of the strongest ones smaller than 5$\AA$ \citep{1991ApJ...374..431S,1991ApJS...76..813S}.

Since the discovery of the first extragalactic source in the VHE domain \citep[Mrk~421, ][]{1992Natur.358..477P}, the number of detected objects increased from a few in early 2000 up to 50 objects at the beginning of 2013 (see
TeVCat\footnote{\url{http://tevcat.uchicago.edu}} for an up-to-date overview.). The advent of the current generation of atmospheric Cherenkov
telescopes (\hess, VERITAS, MAGIC) and the subsequent  gain in sensitivity permit the detection of fainter and more distant sources. The efforts spent on understanding the properties of blazars and predicting VHE emission \citep{REF::TEVCANDITATE,2008A&A...478..395M,2010ApJ...716...30A}, together with measurements in the high-energy (HE; 100~MeV$<$E$<$300~GeV) range, greatly contribute to the detection of new sources.

The spectral energy distribution (SED) of blazars is bimodal with one
bump at low energy (from radio up to X-rays) and one at higher energy (from
X-rays to TeV). BL~Lac objects are further divided into two subclasses depending
on the ratio between the X-ray and the radio fluxes \citep{1995ApJ...444..567P}: the
low-frequency peaked BL~Lac objects with a peak below UV wavelengths and
the high-frequency peaked BL~Lac (HBL) objects for which the peak is in the UV
or X-ray range. The latter subclass represents the bulk of the currently known
extragalactic VHE $\gamma$-ray emitters detected by atmospheric Cherenkov
telescopes (34 out of 50 early 2013), and almost 50\% of the
second \fer\ catalogue of HE AGN \citep[2LAC,][]{2011ApJ...743..171A}.

The low-energy bump of the SED is attributed to synchrotron emission of
relativistic leptons (e$^+$e$^-$ pairs) moving along the jet. The origin of the high-energy
component is less definite. Leptonic models invoke inverse Compton (IC) scattering
on either the synchrotron photons \citep[synchrotron self-Compton, e.g.
][]{THEO::SSC_BAND} or an external photon field \citep[external Compton, e.g.
][]{Dermer1993}. The \grs\ can also be produced by hadronic interactions, such as
photo-production of pions \citep[e.g. ][]{1993A&A...269...67M} or synchrotron
emission of protons \citep[e.g. ][]{2000NewA....5..377A}.

First detected in X-rays with ROSAT \citep[1RXS J001356.6--185408,][]{RASS},
\SHBL was later identified as a BL~Lac object by \citet{RBS}. With a radio flux
of 29.6 mJy at 1.4 GHz \citep{NVSS} and an X-ray flux, between 0.1--2.4 keV, of $1.26 \times 
10^{-11}$\ergcms\ \citep{RASS}, it fulfilled the selection criteria of the sedentary survey of
extreme high-energy peaked BL~Lacs \citep[SHBL,][]{2005A&A...434..385G}, making
this source a member of the HBL subclass.

Its relative proximity \citep[z=0.095,][]{2009MNRAS.399..683J} and its X-ray
and radio fluxes are criteria that make \SHBL an interesting target for very
high-energy (VHE, $E>100\,{\rm GeV}$) observations \citep{REF::TEVCANDITATE}.
Consequently, the source was observed with the High Energy Stereoscopic System
(\hess) and indeed reported as a TeV \gr\ emitter in November 2010
\citep{2010ATel.3007....1H}.

The \fer\ Large Area Telescope (LAT), launched on June 11, 2008, did not detect
a high-energy (HE; 100~MeV$<$E$<$300~GeV) counterpart after two years of
operation \citep[\fer\ two-year catalogue, 2FGL,][]{2012ApJS..199...31N}.
However, the analysis of 3.5 years of data reported here reveals the presence of
a faint source that is positionally coincident with \SHBLc.

The \hess\ and \fla\ data analyses and results are presented in sections
\ref{hess} and \ref{fermi}. The multi-wavelength data set is presented in
sections \ref{swiftxrt} and \ref{swiftuvot} for the X-ray and UV observations with \swift\ and in section
\ref{atom} for the optical observation with ATOM. The discussion in section \ref{SED} focusses on the
description of the SED in the framework of a synchrotron self-Compton (SSC)
model.

Throughout this paper a $\Lambda$CDM cosmology with H$_0$ = 71 km s$^{-1}$
Mpc$^{-1}$, $\Omega_\Lambda$ = 0.73 and $\Omega_{\rm M}$ = 0.27 is assumed,
resulting in a luminosity distance of D$_{\rm L}$ = 431 Mpc \citep{1999astro.ph..5116H,2006PASP..118.1711W}.

\section{Observations and analyses}\label{ObsMeth}
\subsection{\hess\ data set and analysis} \label{hess}

\hess is located in the Khomas Highland, Namibia (23\dgr16'18'' S, 16\dgr30'01''
E), at an altitude of 1800 m above sea level. \hess\ is an array of imaging
atmospheric Cherenkov telescopes. Each of four telescopes used in this study (\hess\ phase 1) consists of a segmented 13 m
diameter optical reflector \citep{Bernlohr} and a camera composed of 960
photomultipliers. The system works in a coincidence mode requiring the detection
of an air shower by at least two telescopes \citep{Funk}.

\SHBL was observed with \hess\ between MJD~54653 and MJD~55912 (July, 6 2008
-- December, 17 2011). Data were selected using the standard quality
criteria \citep[good weather, no hardware problem, see][]{aha2006}, yielding an
exposure of $41.5\,{\rm h}$ acceptance corrected live time at a mean zenith
angle of $12.9^\circ$. The {\it Model} analysis \citep{Naurois} was
performed with the {\it standard cuts} (threshold of 60 photo-electrons),
leading to an energy threshold of $E_{\rm th}=310\,{\rm GeV}$ for this observation.
An excess of 153 \gr\ candidates has been found using the
\textsl{Reflected-background} method \citep{aha2006} to subtract the background.
The total numbers of ON- and OFF-source events are $\rm N_{ON} = 830$ and $\rm
N_{OFF} = 8190$, respectively, with a background normalization factor $\alpha
\simeq 0.083$. The source is detected with a significance of $5.5\,\sigma$
\citep[Eq. 17 of ][]{Lima}. 

\begin{figure}[tbh]
\centering
\includegraphics[width=0.8 \linewidth]{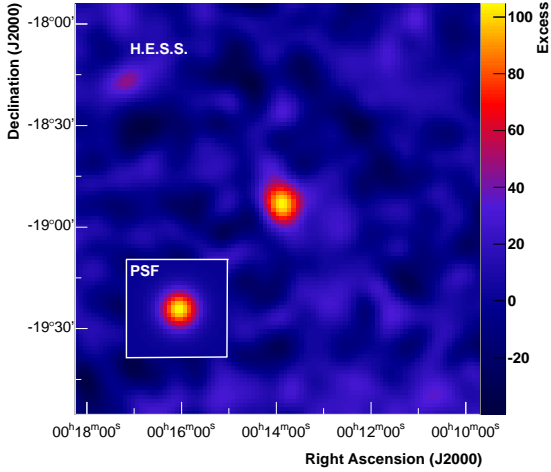}
\caption{Map of the \gr\ excess measured with the \hess\ telescopes around the position of
\SHBL in right
ascension and declination (J2000). The map is smoothed with the \hess\ PSF, shown in inset.}
\label{fig:map}
\end{figure}

\begin{figure}[tbh]
\centering
\includegraphics[width=0.99 \linewidth]{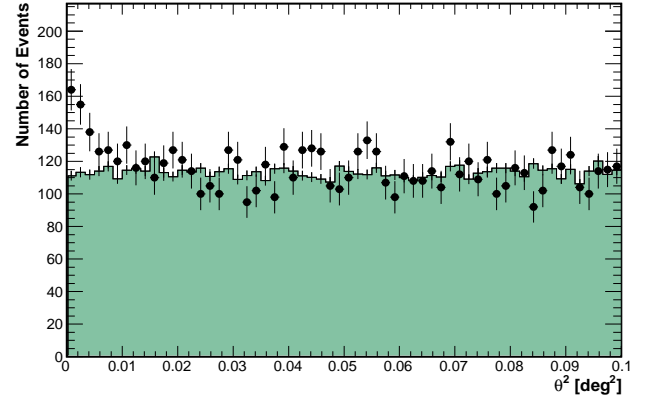}
\caption{Distribution of ON-source events (circles) as a function of the
 square of the angular distance $\theta^2$ from the source position. A cut at $\theta=0.1\dgr$ is used to define the ON region with {\it standard
cuts}. The distribution of normalized OFF-source (solid histogram) is flat and matches the distribution of the ON-source events at large $\theta^2$, which reflects a proper background subtraction.}
\label{fig:theta}
\end{figure}

The distribution of excess events in a 2D map of 2\dgr\ field of view centred on
the source coordinates is given in Fig.~\ref{fig:map}. The fit of a point-like
source, convolved with the \hess\ point spread function (PSF), to the data
results in the best-fit position of $\alpha_{\rm J2000}= 00^{\rm h}\,13^{\rm
m}\,52^{\rm s}\pm 1.5^{\rm s}_{\rm stat}\pm 1.3^{\rm s}_{\rm sys}$ and
$\delta_{\rm J2000}=-18^\circ\,53'\,29\arcsec\pm 22\arcsec_{\rm stat} \pm 20\arcsec_{\rm sys}$,
systematic uncertainties arising from telescope pointing. This is less than $2
\sigma_{\rm stat}$ away from the test position of \SHBL, derived from radio
observations \citep[$\alpha_{\rm J2000}= 00^{\rm h}\,13^{\rm m}\,56^{\rm s} \pm
0.05^{\rm s}$ and $\delta_{\rm J2000}=-18^\circ\,54'\,06\arcsec \pm 0.7\arcsec$,
][]{NVSS}. The source is compatible with a point-like source within the
systematic uncertainties on the \hess\ PSF. The distribution of ON-source events
as a function of the square of the angular distance to the radio position is
shown in Fig.~\ref{fig:theta} along with the OFF-source distribution.

\begin{figure}[tbh]
\centering
\includegraphics[width=0.99 \linewidth]{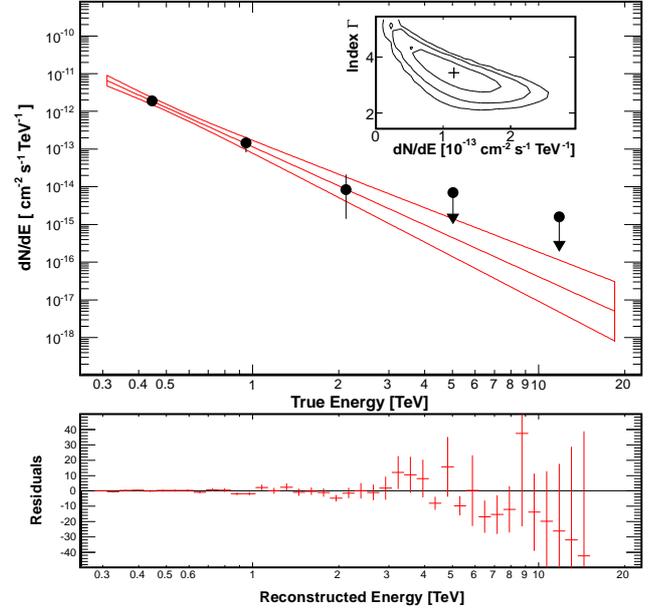}
\caption{Spectrum of \SHBL measured with \hess\ using the {\it Model} analysis.
Top panel: The butterfly represents the $1\,\sigma$ contour for the best-fit model. The spectrum is obtained with the forward-folding method in \citet{2001A&A...374..895P}. The data points are derived {\it a posteriori} and should be considered as residuals. For bins with a significance below $2\sigma$,  upper limits at the 95\% confidence level are computed. The inset gives the 1, 2, and $3\,\sigma$
confidence levels in the power-law index vs differential flux at 1~TeV plane. Bottom panel: Residuals of the fit, with the binning used in the forward-folding method.}
\label{fig:spec}
\end{figure}

The VHE spectrum (Fig.~\ref{fig:spec}), derived from the \hess\ data using the
forward-folding method in \citet{2001A&A...374..895P}, is compatible\footnote{The
$\chi^2/{\rm NDF}$ is $25.1/19$ for a corresponding probability of $15\%$.} with a
simple power law of the form
\[
\frac{{\rm d}N}{{\rm d}E}=(\hessfluxEdec) \times
  10^{-12}\left( \frac{E}{E_{\rm dec}}
  \right)^{-\indexHESSnumOnly} \mathrm{cm}^{-2}\,\mathrm{s}^{-1}\,\mathrm{TeV}^{-1},\]
where the decorrelation energy is $E_{\rm dec} = 510\,{\mathrm{GeV}}$. The
integrated flux above 310~GeV is $(8.3 \pm 1.7_{\rm{stat}}\pm
1.7_{\rm{sys}})\times 10^{-13}$ photons \cms. The $1\,\sigma$ error contour was computed using Eq.~1 in \citet{2010ApJ...708.1310A}, and the systematic
errors were evaluated following \citet{aha2006}. The analysis was cross-checked with an independent
method \citep{Becherini} giving compatible results.

A search for variability was performed on a period\footnote{A period is
defined as the time between two full moons (lunation).} time scale. The corresponding
light-curve (Fig.~\ref{fig:lc}, top panel) does not show any significant variations exceeding the
experimental uncertainties, and the fit with a constant function yields a $\chi^2$
probability of 7\%. The fractional excess variance $\fvar$ \citep{Vaughan} is compatible
with zero at the $2\,\sigma$ level, and the 99\% confidence level upper-limit is
$\fvar< 2.57$.

\subsection{\fla\ data set and analysis} \label{fermi}
The LAT on-board the \fer\ satellite is a pair conversion
telescope sensitive to \grs\ between 20 MeV and 300 GeV. Its main
characteristics and performance can be found in \citet{2009ApJ...697.1071A}.
The bulk of LAT observations are performed in an all-sky survey mode, where all objects are seen for about 30 minutes every 3 hours.

Events passing the {\tt SOURCE} selection \citep{2012ApJS..203....4A} with a reconstructed energy between
300~MeV and 300~GeV have been considered in this analysis. The corresponding
instrumental response functions (IRFs) {\tt P7SOURCE\_V6} and the latest public
 version of the ScienceTools (v9r27p1), available from the \fer\ 
Science Support Center\footnote{\url{http://fermi.gsfc.nasa.gov/ssc/data/analysis/}}
web site, were used. A region of interest (ROI) of 15\dgr\ radius around the radio
coordinates of \SHBL was defined to perform a binned analysis
\citep{mat96}, implemented in the {\tt gtlike} tool. Additionally, cuts were
applied on the rocking angle of the spacecraft, which was required to be
smaller than 52\dgr, and on the zenith angle of the events, required to be smaller than 100\dgr.

The model used to describe the observed emission consists of all the sources
contained in the 2FGL catalogue within this ROI, an isotropic extragalactic
diffuse component described with the file {\tt iso$\_$p7v6source.txt}, and the
standard Galactic model {\tt gal$\_$2yearp7v6$\_$v0.fits}. At the position of \SHBLc, a point-like source with a
power-law model was added. The model of \SHBL has two free parameters: the
differential flux at a fixed energy and the photon index $\Gamma$. The spectral
parameters of sources within 3\dgr\ of the \SHBL position, as well as the
diffuse background normalizations and the variable and/or bright sources (namely
2FGL J2330.9--2144 and 2FGL J2345.0--1553), were left free to vary during the
fitting procedure, whereas the other parameters were frozen to the 2FGL values.

Using 1290 days ($\simeq 3.5$ years) of data, from August, 4 2008 to February, 12
2012 (MJD~54682 to MJD~55973), \SHBL is detected with a test statistic\footnote{see \url{http://fermi.gsfc.nasa.gov/ssc/data/analysis/documentation/Cicerone/Cicerone_Likelihood/Likelihood_overview.html} for a definition.} (TS) of
34, which approximately corresponds to $5.5\,\sigma$. This source was not detected by \fla\ at the time of the 2FGL catalogue, which can be explained by its faint \gr\ flux. The coordinates derived
with {\tt gtfindsrc} are compatible with the radio position within $1\,\sigma$.
The source spectrum is described well by a power law with a differential flux of
$(3.5\pm0.9_{\rm stat}\pm0.7_{\rm
sys})\times10^{-14}\,\mathrm{cm}^{-2}\,\mathrm{s}^{-1}\,\mathrm{MeV}^{-1}$ at
the decorrelation energy $E_{\rm dec}= 2.90$ GeV. The photon index of the source
is found to be $\Gamma = 1.96 \pm 0.20_{\rm stat} \pm 0.08_{\rm sys}$, yielding a
total flux above 300~MeV of ($9.3 \pm 3.4_{\rm stat} \pm 0.8_{\rm sys})\times
10^{-10}$ photons \cms. The photon with the highest detected energy that is very likely
associated with the source\footnote{A photon is considered as associated to the
source if its reconstructed direction is within the 95\% containment radius of
the \fla\ PSF.} has an energy $E\approx37$~GeV. Systematic uncertainties have
been evaluated using the IRFs bracketing method \citep{2009ApJ...707.1310A}. A
fit with a log-parabola function does not improve the overall results. The
differential flux has also been measured in four energy bins (see
Fig.~\ref{fig:spec}) by performing a {\tt gtlike} analysis for which the photon
index of the source is frozen to the best-fit value. The last bin has a TS$<$8
and a 95\% confidence level upper limit has been computed.

Due to the faint emission of the source in the 300~MeV--300~GeV range, a light-curve
with six-months-wide time-bins was computed using the {\tt gtlike} analysis
chain. This light curve is presented in the second panel of Fig.~\ref{fig:lc}.
Upper limits at the 95\% confidence level were calculated for time bins with TS$<$4. No significant
variability can be measured on this time scale with $\fvar<1.83$.

\subsection{\swiftxrt\ data set and analysis} \label{swiftxrt}

X-ray observations are usually an important part of characterizing the
SED of blazars, since they probe the synchrotron radiation of the highest energy
leptons. Target-of-opportunity observations with the space-based \swift\
\citep{2005SSRv..120..165B} X-ray observatory have been conducted
in September 2010. Data taken in photon-counting (PC) mode are processed with
the standard {\em xrtpipeline} tool from the {\tt HEASOFT V6.12}, where a King
function fit to the PSF shows no evidence of any pile-up in any of the four
observations. Events and background extraction regions are defined with a 60-pixel radius circle (corresponding to $\approx 142$\arcsec), with the latter
centred near the former without overlapping. The \swiftxrt\ spectrum has been
rebinned so as to have at least 20 counts per bin using {\tt grppha}, yielding
a usable energy range between $0.3$ and $9.0\,{\rm keV}$ for the summed spectrum
( $0.3$ and $ 5-7\,{\rm keV}$ for single observations). The weighted average
column density of Galactic HI $N_{\rm H}=2.13\times10^{20}\,{\rm cm^{-2}}$ has
been extracted from the Leiden/Argentine/Bonn (LAB) Survey \citep{Kalberla2005}
using the nH tool from {\em
HEASARC}\footnote{\url{http://heasarc.gsfc.nasa.gov/cgi-bin/Tools/w3nh/w3nh.pl}}.
Multiple model spectra are tested with {\tt pyXspec 1.0}, using the response
functions {\tt swxpc0to12s6\_20010101v013} and a dedicated ancillary response
function (ARF) using {\tt xrtmkarf} within {\tt FTOOLS} at the location of the
source in the field of view and with the summed exposure maps of the single
observations. A single power law $F(E)=K E^{-\Gamma_X}$ poorly represents the
fitted summed spectra, with $\chi^2_r=1.23(179)$. An F-test probability
\citep{bevington} of $5\times10^{-4}$ prefers the simplest smooth curved
function, a three-parameter log-parabola $F(E)= K E^{-a - b\log(E)}$,
for which all parameters are given in Table~\ref{table:xrtfit}, including the
spectrum obtained after summing all four event files in {\tt xselect V2.4b} and
building an exposure map with {\tt ximage V4.5.1}. Similar conclusions were drawn by \citet{2011ApJ...739...73M} on this object with a different calibration. The source shows no
indication of variability over the span of eight days (third panel of
Fig.~\ref{fig:lc}).

\begin{table*}
\caption{Parameters of power-law or log-parabolic fits to all four \swiftxrt\ observations, as well as the
  summed spectrum. Column 4 gives either the power-law index $\Gamma_X$ or the log-parabola value of $a$ depending on which model best fits the data. Column 7 shows the reduced $\chi^2$ value $\chi^2_r$,
  column 8 the F-test probability for the log-parabolic model when it is preferred over the power law, and column
9 the estimation for the unabsorbed flux (using {\tt cflux}) in the $0.3-10\,{\rm keV}$ range. The values of $b$ and
F-test are not provided when the log-parabola is not preferred. } 
\label{table:xrtfit}
\centering
\begin{tabular}{c c c c c c c c c} 
\hline\hline 
ObsID & Start time & exposure & $\Gamma_X$ or $a$ & $b$ & $K$ & $\chi^2_r$(ndf) & F-test & $F_{0.3-10\,{\rm keV}}$\\
 & MJD-55000 & [s] &  &  & [$10^{-3}\,{\rm cm^{-2}}\,{\rm s^{-1}}\,{\rm keV^{-1}}$] & & & [$10^{-11}\,{\rm erg}\,{\rm cm^{-2}}\,{\rm s^{-1}}$]\\   
\hline 
00031806002 & 449.05 & 3783 & $1.95\pm0.06$ & $0.30^{+0.14}_{-0.13}$ & $2.62\pm0.09$ & $1.15(55)$ &  $4\times10^{-2}$ & $1.28\pm0.06$\\
00031806003 & 449.98 & 4190 & $2.03\pm0.04$ &  & $2.46\pm0.07$ & $1.28(61)$ &  & $1.38\pm0.04$\\
00031806004 & 451.05 & 3848 & $1.98\pm0.05$ & $0.26_{-0.13}^{+0.14}$ & $2.46\pm0.09$ & $0.98(52)$ & $5\times10^{-2}$ & $1.30\pm0.08$\\
00031806005 & 451.58 & 2745 & $2.02\pm0.05$ &  &  $2.38\pm0.08$ & $0.80(39)$ &  &$1.33\pm0.09$\\
sum && 14566 & $2.00\pm0.03$ & $0.23\pm0.06$ & $2.73\pm0.05$ & $1.15(178)$ & $5\times10^{-4}$ & $1.36\pm0.03$ \\

\hline 
\end{tabular}
\end{table*}

\subsection{\swiftuvot\ data set and analysis} \label{swiftuvot}

The UVOT instrument \citep{Roming2005} on-board {\it Swift} measured the UV
emission of \SHBL in the bands U (345 nm) and UVW2 (188 nm) simultaneously with
the X-ray telescope with an exposure of $\sim 1.5 - 2\;\rm{ks}$ each. The
instrumental magnitudes and the corresponding fluxes (for conversion factors, see
\citealt{Poole2008}) are calculated with {\tt uvotmaghist} taking all photons into account from a circular region with a radius of $5''$ (standard aperture for
all filters). An appropriate background was determined from a circular region
with radius $40''$ near the source region without contamination of neighbouring
sources. The measured UV fluxes are corrected for dust absorption using $E(B-V)
= 0.0246$ mag\footnote{\url{http://irsa.ipac.caltech.edu/applications/DUST/}}
(Schlegel et al. 1998) and the A$_\lambda/E(B-V)$ ratios given in
\citet{1979MNRAS.187P..73S}.

The flux (Fig.~\ref{fig:lc}, fourth panel), binned per observations, does not show any sign of
variability in either of the two bands. Table~\ref{table:atom} gives the chance
probability obtained when fitting the data with a constant, as well as the
average flux obtained when analysing all the observations together. 

\subsection{ATOM data set and analysis}
\label{atom}

The 75~cm Automatic Telescope for Optical Monitoring \citep[ATOM, ][]{ATOM},
located on the \hess\, site has been used to monitor the optical emission of
\SHBL in Bessell B, R, and I filter bands \citep{Bessel90} over the last five years.
The presented data have been obtained from MJD~54629 to MJD~55897 (June, 6 2008
-- December, 2 2011). A total of 138, 188, and 2 observations in B, R,
and I bands, respectively, were carried out with an aperture of 4$\arcsec$ radius. The data were
corrected for Galactic absorption using A$_\lambda(B) = 0.107$ mag,
A$_\lambda(R) = 0.066$ mag and A$_\lambda(I) = 0.048$ mag \citep{Schlegel} and
the Bessell zero points \citep{Bessel90} are used to convert the magnitude into
flux units. The time-averaged flux in each band is given in Table
\ref{table:atom} along with the corresponding energy.

\begin{figure}[tbh]
\centering
\includegraphics[width=0.99 \linewidth]{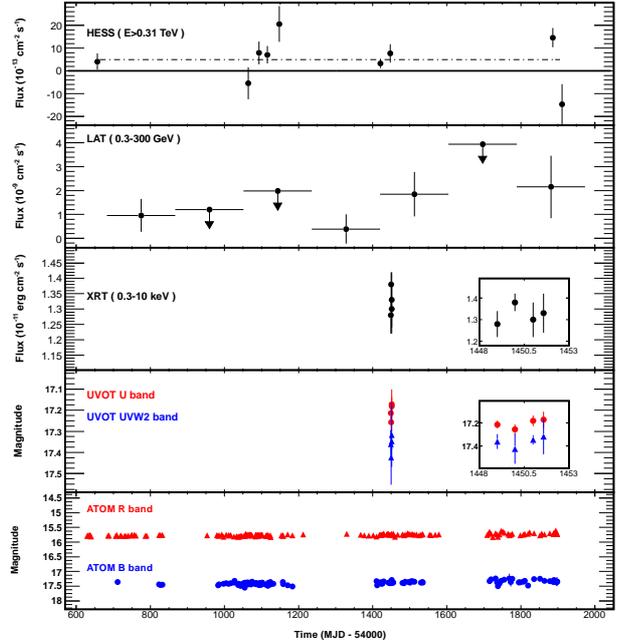}
\caption{Light-curves from observations at different wavelengths. From top
to bottom: {\em i}) \hess\ period-by-period integral-flux above 310~GeV. The dashed line is the mean flux level. The
vertical bars give the statistical errors. {\em ii}) \fla\ flux in the
300~MeV--300~GeV energy range. For time bins where the test-statistic is below 4,
an upper limit at 95\% confidence level is reported. {\em iii}) \swiftxrt\ flux in the 0.3--10~keV band. {\em iv})
\swiftuvot\ measurement in U and UVW2 bands. {\em v}) ATOM magnitudes in the B and R bands. The insets of panel {\em iii} and {\em iv} present a zoom in time around the \textit{Swift} observations.}
\label{fig:lc}
\end{figure}

Table~\ref{table:atom} also gives the $\chi^2$ probability obtained when fitting
the time series with a constant function. The flux of the source
(Fig.~\ref{fig:lc}, fifth panel) is compatible with being constant over time in
each band.

\begin{table}
\caption{\swiftuvot\ U and UVW2 measurements and time-averaged optical magnitudes and fluxes measured with ATOM in
R, B, and I Bessel filters. The last column gives the probability obtained when fitting the time series with a constant function. Magnitudes are not corrected for Galactic extinction, whereas fluxes are dereddened using the values of \citet{Schlegel}.}
\label{table:atom} 
\centering 
\begin{tabular}{c c c c c} 
\hline\hline 
 Filter & Energy  & magnitude & Flux  & P($\chi^2_r$)  \\ 
& [eV] & & [$10^{-12}\ \ergcms$] & \% \\
\hline 
UVW2 & 6.6 & 17.4 & 1.99 & 63 \\
U & 3.6 & 17.2 & 1.44 & 94 \\
\hline 
B & 2.88 & 17.5 & 3.26 & 99  \\
R & 1.77 & 15.8 & 6.09 & 19  \\
I & 1.37 & 14.2 &  8.36 &   \\
\hline 
\end{tabular}
\end{table}
\section{Discussion}
\label{SED}

\begin{figure}[tbh]
\centering
\includegraphics[width=0.99 \linewidth]{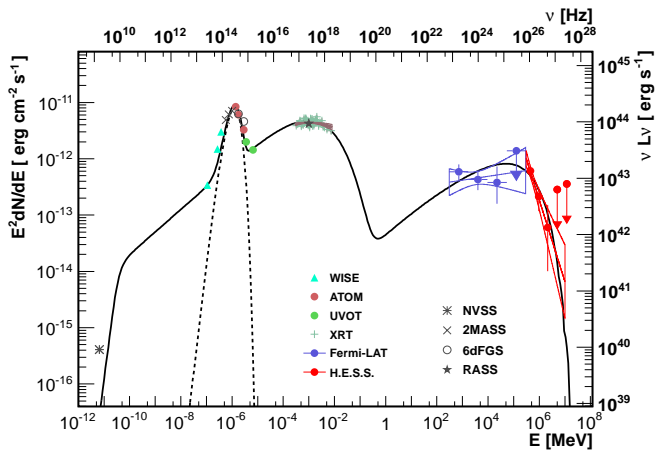}
\caption{Radio-to-TeV spectral energy distribution of \SHBL with ATOM (red
circles, Table~\ref{table:atom}), \swift-UVOT (green circles), \swiftxrt\ (green cross + brown lines, Table~\ref{table:xrtfit}),
\fla\ (blue), and \hess\ (red) measurements. Archival data from NVSS, 2MASS,
6dFGS, and RASS are shown in grey. The dashed line is the black-body spectrum from the
host galaxy at a temperature of 4500~K, and the black line shows the sum of the 
SSC calculation and the black-body spectrum.}
\label{fig:sed}
\end{figure}

The broadband data presented in this work from ATOM, \swift, \fer, and \hess,
together with archival data from the NVSS \citep{NVSS}, 2MASS \citep{2MASS},
6dFGs \citep{2009MNRAS.399..683J}, and RASS \citep{RASS} catalogues are
used to constrain physical characteristics of the source. Contemporaneous data from the {\em Wide-field Infrared Survey Explorer} \citep[{\em WISE}, ][]{WISE} in the bands 3.4, 4.6, and 12 $\mu$m are also presented. The SED exhibits the usual
two components, as can be noticed from Fig.~\ref{fig:sed}.

The \swiftxrt\ data allow the synchrotron peak position to be located at $E_{\rm
s,\,peak} = 1.00\pm0.07$~keV, a quite common feature of many objects in the
SHBL catalogue, that peak in X-ray as mentioned by
\citet{2005A&A...434..385G}. Unfortunately, the source is too faint for the
energy of the IC peak to be measured accurately, but \citet{2010ApJ...716...30A}
found an empirical relation between the \fer\ photon index and the peak position
of the IC component. Applying the relation to \SHBL leads to $E_{\rm
ic,\,peak}=3.1^{+16.3}_{-2.7}\,{\rm GeV}$\footnote{The use of a relation derived
on a population of sources might lead to large uncertainties but in the case of
\SHBL, a factor 10 in $E_{\rm ic,\,peak}$ would not affect the conclusions of this work.}.

The spectral index found between the optical and UV waveband $\alpha_{\rm o-uv}$
is $\simeq -2.3$ which is not in agreement with the one observed between UV and X-ray,
$\alpha_{\rm uv-x} \simeq 1.5$. The negative value of the former indicates
that it does not originate from synchrotron emission of non-thermal particles 
but rather from the host galaxy thermal emission. This is supported by the
non-variability of the ATOM light-curves and the good agreement of the optical
flux with the archival data. Placing the source in a [3.4]-[4.6]-[12] $\mu$m colour-colour diagram \citep{2011ApJ...740L..48M} reveals that the infrared measurement is also dominated by this thermal emission. The host galaxy emission is modelled using a
black-body, as in \citet{Katarz03}, with a temperature of 4500 K and a total
luminosity of $1.43\times 10^{44}$ \ergs, which corresponds to $3.7\times 10^{10}$ times the luminosity of the Sun.

The non-thermal emission is modelled with a simple, time-independent, one-zone homogeneous
SSC model \citep{THEO::SSC_BAND}. The emission zone
is spherical with a radius $R$, moving relativistically towards the Earth with a Doppler factor $\delta$. A
population of leptons with a density $N_l(\gamma)$ is responsible for the
synchrotron emission by interacting with a uniform magnetic field $B$ and for
the \gr\ emission by IC scattering of the synchrotron photons.

Some properties of the SSC model can be assessed using the available data. A
constraint on the maximal energy of the leptons can be found assuming the last
bin in energy of \swiftxrt\ is a lower limit for the maximum energy reached by the synchrotron
process $E_{\rm s,\,max}$. This gives \citep{BOOK:RandB79}

\begin{equation}
E_{\rm s,\,max} \approx \frac{21\delta \hbar
  \gamma_{\rm max}^2qB}{15(1+z)\sqrt{3}m_e} \gtrsim 8\,{\rm keV},
\label{eq:1}
\end{equation}
where $q$ is the charge of the electron and $B$ is in Gauss. Equation \ref{eq:1} implies $\gamma_{\rm max} \geq 9.7\times 10^4 \ B^{-1/2}\delta^{-1/2}$.

Since the same population of particles is responsible for the two spectral
components and under the hypothesis that the IC scattering occurs in the Thomson
regime, Eq. 4 of \citet{1998ApJ...509..608T} reads as
\begin{equation}
B\delta = (1+z) \frac{8.6\times10^7E_{s,\rm peak }^{2}}{E_{\rm ic,\,peak}} =
3.1\times 10^4\ {\rm G},
\end{equation}
where the energies are expressed in eV. For a magnetic field $B$ of
0.1~G, this leads to a Doppler factor of $3.1\times 10^5$, which is unrealistic.
For the photons of energy higher than $E_{\rm ic,\,peak}$, the scattering must
occur in the Klein-Nishina regime in which case the leptons producing the
highest energy detected follow $E_{\rm ic,\,max}=\frac{\delta}{1+z}\gamma_{\rm
max}m_ec^2=18\,{\rm TeV}$, which with the above constraint on $\gamma_{\rm
max}$, in turn yields $B\delta^{-1} \geq 6.3\times 10^{-8}\,{\rm G}$.

Even with the previous calculations, the SSC model is still not fully
constrained by the data. To reduce the number of parameters, the
density of leptons is described by a power law with an exponential cut-off of
the form $N_e(\gamma)\propto\gamma^{-p}\cdot\exp(-\gamma/\gamma_{\rm cut})$
which only has three parameters. An equipartition factor $Q$ (defined as the ratio
between the density of particle kinetic energy and the magnetic energy density
(i.e. $Q = u_e/u_B$)), that is as close to unity as possible has also been required, to
minimize the total energy of the jet \citep[see][]{2012arXiv1202.5949G}. With
this model, the minimum factor allowing a fit to the data is $Q=50$, which rules
out the equipartition and the jet is found to be particle-dominated. The same
conclusions have been drawn for other sources, such as Mrk~421
\citep{2011ApJ...736..131A} or Mrk~501 \citep{2011ApJ...727..129A}.

The SSC calculation, together with the thermal black-body component, is shown in Fig.~\ref{fig:sed}, and parameters are given in
Table~\ref{table:ssc}. The attenuation by the extragalactic background light has
been taken into account using the model of \citet{2008A&A...487..837F}. The
index of the leptonic distribution is found to be $p=2.2$ with a break at
$\gamma_{\rm cut} = 5.0\times 10^5$, for a total number of leptons of $2.0\times
10^{53}$. The size of the emission region is $R=3.5\times 10^{16}$ cm, the
Doppler factor $\delta=10$, and the magnetic field $B= 0.05$ G (i.e. $U_B =
10^{-4}\ \mathrm{erg}\ \mathrm{cm}^{-3}$), values that comply with the above-mentioned 
constraints. The minimum variability time scale achievable
within this scenario is 1.4 days, which cannot be tested with \hess\ or \fer\
given the flux of the source.

To check that $\gamma_{\rm cut}$ can be related to the synchrotron
cooling time or be acceleration effects, the comparison between the ratio of
the light crossing time $\tau_c$ in the emission zone rest-frame and the cooling
time\footnote{The inverse of the cooling time for synchrotron dominated models
is defined as $\tau_{\rm cool}^{-1} = {4 \over 3}\ {{\sigma_T c} \over {m_e
c^2}} \ \gamma_{\rm cut} \ u_{B}$.} $\tau_{\rm cool}$ can be used.
\citet{1998ApJ...509..608T} suggest that this ratio should be between 1 and 3.
For the model presented here, this ratio is close to 1.8 and so is not possible
to disentangle a radiative cooling break from a cut-off in the lepton distribution due to acceleration effects.

\begin{table}
\caption{SSC parameters used to reproduce the SED of \SHBLc. The calculation is shown as a black line in Fig.\,\ref{fig:sed}}
\label{table:ssc} 
\centering 
\begin{tabular}{c  c } 
\hline\hline 
Parameters & Value     \\ 
\hline 
$B$ [G] & $0.05$  \\
$R$ [cm] & $3.5\times 10^{16}$   \\
$\delta$ & $10$  \\
\hline 
$p$ & $2.2$  \\
$\gamma_{\rm min}$ & 1\\
$\gamma_{\rm cut}$ & $5.0\times 10^5$\\
$N_{\rm tot}$ & $2.0\times 10^{53}$ \\
\hline 
\end{tabular}
\end{table}

\section{Conclusions}

Dedicated observations using the \hess\ telescopes have revealed a new HBL,
\SHBLc, as a VHE \gr\ emitter, with a significance of $5.5\,\sigma$. The source has a
flux of about 0.6\% of the Crab Nebula flux above 310~GeV, with a soft photon index \indexHESS. 

Using 3.5 years of \fer\ data, the presence of a previously undetected
counterpart in the HE range was found. Both HE and VHE spectra connect
smoothly, leading to the conclusion that the same population of particles is likely to be 
responsible for the \gr\ emission from 300 MeV to a few TeV.

At lower energy, the optical measurements, which are contemporaneous with the \hess\ and
\fla\ observations, are found to be constant over time and compatible with the
data taken over the past decade. They are interpreted as thermal emission of the
host galaxy and successfully reproduced by a black-body model.

An SSC model has been used to reproduce the non-thermal emission from radio to TeV
energies. In this model, the \grs\ with energies above the \gr\ peak are produced by IC 
scattering showing evidence for Klein-Nishina suppression. Equipartition between the
kinetic and the magnetic energy density is ruled out by the calculation, implying that the jet
is dominated by the lepton kinetic energy.

The number of extragalactic sources jointly detected by the current generation
of atmospheric Cherenkov telescopes and the \fla\ is increasing rapidly, allowing the
\gr\ sky from 100\, MeV to a few TeV to be probed and the mechanisms responsible
for the electromagnetic emission to be constrained. The knowledge of the
non-thermal sky will increase with the advent of \hess\ 2 and the future
Cherenkov Telescope Array (CTA). The ten-fold increased sensitivity of CTA with respect to current generation atmospheric Cherenkov telescopes and the possibility of performing an extragalactic survey \citep{2012arXiv1208.5686D} will allow detection of hundreds of sources with fluxes of 1\% of that the Crab Nebula. Such survey, together with ten years of \fla\ data, will allow detailed population studies.

\acknowledgements
The support of the Namibian authorities and of the University of
Namibia in facilitating the construction and operation of H.E.S.S.
is gratefully acknowledged, as is the support by the German
Ministry for Education and Research (BMBF), the Max Planck
Society, the French Ministry for Research, the CNRS-IN2P3 and the
Astroparticle Interdisciplinary Programme of the CNRS, the U.K.
Particle Physics and Astronomy Research Council (PPARC), the IPNP
of the Charles University, the South African Department of Science
and Technology and National Research Foundation, and by the
University of Namibia. We appreciate the excellent work of the
technical support staff in Berlin, Durham, Hamburg, Heidelberg,
Palaiseau, Paris, Saclay, and in Namibia in the construction and
operation of the equipment. 

The \textit{Fermi} LAT Collaboration acknowledges generous ongoing support
from a number of agencies and institutes that have supported both the
development and the operation of the LAT as well as scientific data analysis.
These include the National Aeronautics and Space Administration and the
Department of Energy in the United States, the Commissariat \`a l'Energie Atomique
and the Centre National de la Recherche Scientifique / Institut National de Physique
Nucl\'eaire et de Physique des Particules in France, the Agenzia Spaziale Italiana
and the Istituto Nazionale di Fisica Nucleare in Italy, the Ministry of Education,
Culture, Sports, Science and Technology (MEXT), High Energy Accelerator Research
Organization (KEK) and Japan Aerospace Exploration Agency (JAXA) in Japan, and
the K.~A.~Wallenberg Foundation, the Swedish Research Council and the
Swedish National Space Board in Sweden.

Additional support for science analysis during the operations phase is gratefully
acknowledged from the Istituto Nazionale di Astrofisica in Italy and the Centre National d'\'Etudes Spatiales in France.

This research has made use of the NASA/IPAC Extragalactic Database (NED) 
which is operated by the Jet Propulsion Laboratory, California Institute of Technology, 
under contract with the National Aeronautics and Space Administration.

This research has made use of the VizieR catalogue access tool, CDS, Strasbourg, France.

This publication makes use of data products from the Wide-field Infrared Survey Explorer, which is a joint project of the University of California, Los Angeles, and the Jet Propulsion Laboratory/California Institute of Technology, funded by the National Aeronautics and Space Administration.

The authors want to thank D.~Paneque for the useful comments that improved the paper.

\bibliography{SHBL_detection}
\bibliographystyle{bibtex/aa}
\end{document}